\documentclass[12pt]{iopart}

\usepackage{cite}
\usepackage{graphicx}

\begin{document}

\title{A field theoretic approach to master equations and a variational method beyond the Poisson ansatz}

\author{Jun Ohkubo}

\address{
Institute for Solid State Physics, University of Tokyo, 
Kashiwanoha 5-1-5, Kashiwa, Chiba 277-8581, Japan
}
\ead{ohkubo@issp.u-tokyo.ac.jp}
\begin{abstract}
We develop a variational scheme in a field theoretic approach to a stochastic process.
While various stochastic processes can be expressed using master equations,
in general it is difficult to solve the master equations exactly,
and it is also hard to solve the master equations numerically because of the curse of dimensionality.
The field theoretic approach has been used in order to study such complicated master equations,
and the variational scheme achieves tremendous reduction in the dimensionality of master equations.
For the variational method, only the Poisson ansatz has been used,
in which one restricts the variational function to a Poisson distribution.
Hence, one has dealt with only restricted fluctuation effects.
We develop the variational method further, which enables us to treat an arbitrary variational function.
It is shown that the variational scheme developed gives a quantitatively good approximation 
for master equations which describe a stochastic gene regulatory network.
\end{abstract}

\maketitle

\section{Introduction}

Master equations describe various stochastic phenomena.
For example, a reaction-diffusion process, which is one of the examples of non-equilibrium systems,
is expressed by a master equation.
Usually, it is difficult to obtain the exact solution of the master equation
because of the non-linearity of the corresponding moment equations or its high dimensionality.
The direct numerical solution is also difficult to obtain
because there are an enormous number of coupled equations to be solved.
While numerical simulations such as the Gillespie algorithm \cite{Gillespie1977}
are available for studying complicated stochastic systems,
a coarse-grained analytical approach would be more worthwhile.
The field theoretic approach to the reaction-diffusion process has achieved significant successes\cite{Tauber2005}.
The analogy of the master equation to a quantum system has been introduced by Doi \cite{Doi1976,Doi1976a},
and several authors revived the formalism\cite{Peliti1985,Mattis1998}.
The field theoretic approach has revealed the anomalous kinetics in the reaction-diffusion systems
incorporating the renormalization group method\cite{Tauber2005}.
In addition, 
the field theoretic description has been used 
not only for the reaction-diffusion processes,
but also for various phenomena such as packet flow \cite{Pigorsch2002},
the Malthus-Verhulst process \cite{Dickman2002},
stochastic sandpile models \cite{Dickman2002a,Stilck2004},
and neural networks \cite{Buice2007}.

Recently, Sasai and Wolynes \cite{Sasai2003} have developed the field theoretic approach to a stochastic gene network.
The gene network consists of active and inactive genes, proteins produced by the genes, and 
a mechanism of switching between the active and inactive states caused by the regulatory proteins.
The complicated system is described by a set of master equations, as in the case with 
the reaction-diffusion process.
For the case of only one gene, the exact solution has been obtained \cite{Hornos2005},
but when one consider a general case, i.e., a gene regulatory network,
it is difficult to solve the master equations exactly.
We therefore need an approximation method.
Sasai and Wolynes \cite{Sasai2003} have used the variational method for non-equilibrium systems
which has been proposed by Eyink \cite{Eyink1996,Alexander1996},
The variational method gives us an efficient scheme of approximation for complicated master equations;
it can achieve enormous reduction in the dimensionality of the problem
by solving variationally the quantum field theoretic equations which are obtained by the original master equations.
This means that the variational scheme reduces the coupled master equations with a huge number of variables 
to a set of ordinary differential equations with a small number of parameters.

So far, several schemes of approximation for master equations have been proposed \cite{Risken1984,Gardiner2004}.
In the system size expansion or the Kramers-Moyal expansion \cite{Gardiner2004},
master equation with `discrete' variables are substituted into a Fokker-Planck equation with `continuous' variables.
While the differential equation with continuous variables is easier to treat,
these approximation schemes are applicable only for a system with the large size.
The moment equation approach \cite{Gardiner2004} can be used even for small systems,
and it gives an exact solution when the moment equations are closed.
However, if the moment equations have non-linear terms and are not closed,
it is difficult to obtain the exact solution.
To our knowledge, there is no systematic scheme which produces a closed set of moment equations.
The variational scheme gives a closed set of equations in a systematic way,
and therefore the variational scheme is expected as a candidate 
systematic approximation scheme for complicated master equations.
The variational scheme in \cite{Sasai2003,Kim2007,Kim2007a} is based on the Poisson ansatz,
in which the mean and the variance of the variational function are the same.
It has been revealed that 
the solutions obtained by using the Poisson ansatz are correct only qualitatively
for the repressilator system with two genes \cite{Kim2007,Kim2007a}.

The aim of the present paper is to develop a variational scheme beyond the Poisson ansatz.
In principle, the variational function should be a discrete probability distribution.
The Poisson distribution (the Poisson ansatz) is useful for the functional variation 
because the Poisson ansatz corresponds to the coherent state in the field theoretic description.
On the other hands, the other discrete probability distribution is difficult to treat in the variational scheme.
In order to avoid the difficulty of the variational calculations,
we propose the use of the superposition of the coherent states as the variational function.
By using the superposition of the coherent states,
it becomes possible to assume an arbitrary continuous probability function as the variational function.
We will apply the variational method to a gene regulatory network, which is the same as the one in \cite{Kim2007},
and confirm that the new method gives a quantitatively correct solution.

The construction of the present paper is as follows.
In section~2, we define the gene regulatory network and master equations to be solved,
and we also give the field theoretic description of the master equations.
The variational scheme proposed by Sasai and Wolynes are reviewed in section~3.
Section~4 is the main part of the present paper, and gives the new variational function
beyond the Poisson ansatz.
The numerical experiments are also performed in order to confirm the validity of the new scheme.
Finally, we give some concluding remarks in section~5.

\section{Model and formalism}

\subsection{Master equations of a gene regulatory network}

We here give an explicit example, i.e.,
a chemical reaction network involved in gene regulations, which has been used in \cite{Sasai2003} and \cite{Kim2007}.
The master equations for the gene regulatory network give a closed set of moment equations, 
and hence we can confirm the validity of the variational scheme 
by comparing the results of the variational scheme with those obtained by the moment equations.

\begin{figure}
\begin{center}
  \includegraphics[width=9cm,keepaspectratio,clip]{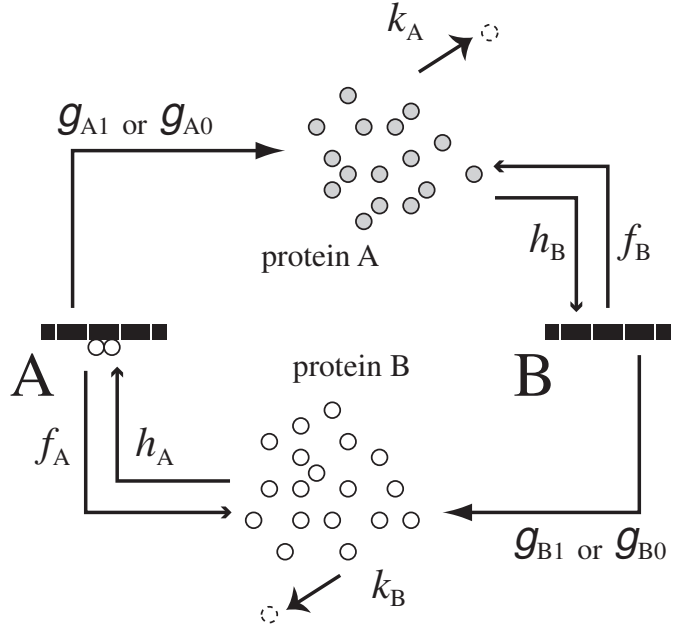} 
\caption{
Illustration of the gene regulatory network.
The Protein produced by gene A is a repressor which binds to gene B, and vice versa.
The production rate $g$ depends on whether the gene is bound or not bound by the repressor.
}
\label{fig_1}
\end{center}
\end{figure}

Figure~\ref{fig_1} shows the gene regulatory network.
In the network, there are two genes which are labeled by A and B, respectively.
Each gene produces a repressor protein which binds to the operator site of the other gene changing the activity.
When gene $\alpha$ ($\alpha = A$ or $B$) is not bound by the repressor proteins,
the gene can produce its own proteins at the rate $g_{\alpha 1}$.
The gene bound by the repressor proteins produces own proteins at the rate $g_{\alpha 0}$.
Note that the subscripts `1' and `0' represent the active state and the inactive state of the gene, respectively.
Each protein spontaneously degrades, and the degradation rate is $k_\alpha$.
The rate of binding of the proteins to a gene and the rate of detaching from a gene are represented by
$h_\alpha$ and $f_\alpha$, respectively.
In the present paper,
we consider the case where dimer proteins repress the expression of a gene.

The next step is to write down master equations for the gene regulatory network.
Hereafter, the number of proteins produced by gene $\alpha$ is denoted as $n_\alpha$.
Using the two component vector notation
\begin{eqnarray}
\mathbf{P}_\alpha (n_\alpha,t) \equiv 
\left(
\begin{array}{c}
P_{\alpha 1}(n_\alpha,t) \\ P_{\alpha 0} (n_\alpha, t)
\end{array}
\right),
\end{eqnarray}
the master equation for the probability with which there are $n_\alpha$ proteins can be written as
\begin{eqnarray}
\fl
\frac{\partial}{\partial t} \mathbf{P}_\alpha (n_\alpha,t)
= \left( \begin{array}{cc}
g_{\alpha 1} & 0 \\ 0 & g_{\alpha 0}
\end{array} \right)
\left[ \mathbf{P}_\alpha(n_\alpha-1,t) - \mathbf{P}_\alpha (n_\alpha,t) \right] \nonumber \\
+ k_\alpha \left[ (n_\alpha + 1) \mathbf{P}_\alpha(n_\alpha+1,t) 
- n_\alpha \mathbf{P}_\alpha (n_\alpha,t) \right] \nonumber \\
+ \left( \begin{array}{cc}
-\frac{h_\alpha}{2} n_\beta (n_\beta -1) & f_\alpha \\
\frac{h_\alpha}{2} n_\beta (n_\beta-1) & -f_\alpha 
\end{array} \right)
 \mathbf{P}_\alpha (n_\alpha,t),
\label{eq_master_equation}
\end{eqnarray}
where $(\alpha,\beta) = \{ (A,B), (B,A) \}$.
Although we might be able to solve the master equation~\eref{eq_master_equation} numerically,
it becomes difficult to solve the master equation numerically when the number of genes increases.
Even for only one gene, we have $2 \times (\textrm{the number of state } n)$ coupled differential equations.
In order to reduce the dimensionality of the problem,
we use the field theoretic description and a variational scheme.

\subsection{Field theoretic description}

It is revealed that the quantum field theoretic method is useful to solve the master equations.
We briefly review the quantum field theoretic description for the gene regulatory network \cite{Sasai2003}.

First of all, we define the ket vector $| n \rangle$ as the state in which there are $n$ proteins in the system.
For each protein (protein A and protein B),
a creation and an annihilation operators are introduced:
\begin{eqnarray}
a_\alpha^\dagger | n_\alpha \rangle = | n_\alpha+1 \rangle,
\end{eqnarray}
\begin{eqnarray}
a_\alpha | n_\alpha \rangle = n_\alpha | n_\alpha-1 \rangle,
\end{eqnarray}
where the index $\alpha$ is $A$ or $B$.
The creation and the annihilation operators satisfy the following commutation relation
\begin{eqnarray}
[ a_\alpha, a_\alpha^\dagger ] = 1,
\end{eqnarray}
and the vacuum state $| 0_\alpha \rangle$ and its conjugate $\langle 0_\alpha |$
are defined to satisfy
\begin{eqnarray}
\langle 0_\alpha | a_\alpha^\dagger = a_\alpha | 0_\alpha \rangle = 0, \\
\langle 0_\alpha | 0_\alpha \rangle = 1.
\end{eqnarray}
Note that the $n$-proteins state $| n \rangle$ is not normalized in the usual sense,
but the states are orthogonal,
because $\langle n | m \rangle = m! \delta_{n,m}$, where $\delta_{n,m}$ is the Kronecker delta,

Using the above quantum field theoretic formalism,
we write the state which corresponds to a probability distribution vector $\mathbf{P}_\alpha (n_\alpha,t)$
as
\begin{eqnarray}
| \psi_\alpha \rangle = \left( \begin{array}{c}
\sum_{n_\alpha} P_{\alpha 1} (n_\alpha,t) | n_\alpha \rangle \\
\sum_{n_\alpha} P_{\alpha 0} (n_\alpha,t) | n_\alpha \rangle 
\end{array} \right).
\end{eqnarray}
The state $| \psi_\alpha \rangle$ only describes the state of gene $\alpha$,
and hence the state of the whole system is denoted by
\begin{eqnarray}
| \Psi \rangle = | \psi_A \rangle \otimes | \psi_B \rangle.
\end{eqnarray}

Next, we introduce the `Hamiltonian' $\Omega$ for the gene regulatory networks.
The Hamiltonian $\Omega$ corresponds to the time-evolution operator in the master equation~\eref{eq_master_equation}.
The master equation~\eref{eq_master_equation} is rewritten in the following form
by using the state defined by $| \Psi \rangle$:
\begin{eqnarray}
\frac{\partial}{\partial t} | \Psi \rangle = \Omega | \Psi \rangle.
\label{eq_master_equation_rev}
\end{eqnarray}
When the total Hamiltonian operator $\Omega$ is defined as
\begin{eqnarray}
\Omega = \Omega_A + \Omega_B,
\end{eqnarray}
the Hamiltonian $\Omega_\alpha$ which operates only gene $\alpha$ is derived from the original master equation as
\begin{eqnarray}
\fl
\Omega_\alpha = 
\left( \begin{array}{cc}
g_{\alpha 1}(a_\alpha^\dagger -1) + k_\alpha(a_\alpha - a_\alpha^\dagger a_\alpha) & 0 \\
0 & g_{\alpha 0}(a_\alpha^\dagger -1) + k_\alpha(a_\alpha - a_\alpha^\dagger a_\alpha)
\end{array} \right)_\alpha \otimes \mathbf{1}_\beta \nonumber \\
+ \left( \begin{array}{cc}
0 & f_\alpha \\
0 &  - f_\alpha
\end{array} \right)_\alpha \otimes \mathbf{1}_\beta 
+ \mathbf{1}_\alpha
\otimes \left( \begin{array}{cc}
\frac{-h_\alpha}{2} (a_\beta^\dagger)^2 (a_\beta)^2  & 0 \\
\frac{h_\alpha}{2} (a_\beta^\dagger)^2 (a_\beta)^2  & 0
\end{array}
\right)_\beta,
\end{eqnarray}
where the suffix $\alpha$ or $\beta$ of each operator means that the operator acts only on gene $\alpha$ or $\beta$.
The first term corresponds to the birth-death part of proteins,
and plays a role in the diffusion effects.
The second and third terms represent the interactions between two genes.
Note that the ``Hamiltonian'' is non-Hermitian and it is a little different from the ordinary quantum mechanics one.
For instances, expected values are linear not bilinear in $|\psi_\alpha \rangle$,
and averages for $|\psi_\alpha \rangle$ are obtained 
by taking the scalar product with the bra vector $( \langle 0 | e^{a_\alpha} \quad \langle 0 | e^{a_\alpha})$.
However, in spite of the non-Hermitian property and a slight difference from the ordinary quantum mechanics,
many quantum field theoretic techniques may be applied, albeit with some modifications.

\section{Variational approach}

\subsection{Variation of the effective action}

In order to reduce the dimensionality of the problem,
a variational method developed by Eyink \cite{Eyink1996,Alexander1996} can be used.
We here briefly review the method \cite{Sasai2003}.

When we define an effective action $\Gamma$ as
\begin{eqnarray}
\Gamma = \int dt \langle \Phi | (\partial_t - \Omega) | \Psi \rangle,
\end{eqnarray}
equation~\eref{eq_master_equation_rev} is equivalent to the functional variation
$\delta \Gamma / \delta \Phi = 0$.
Because of the non-Hermitian property,
it is not always true that the left eigenvectors and right eigenvectors are the same.
Hence, we assume two variational functions for the bra and ket states, respectively.
We assume that the ket state $| \Psi \rangle$ (or the bra state $\langle \Phi |$) 
is parametrized by $\mathbf{x}^R$ (or $\mathbf{x}^L$),
and where $\mathbf{x}^R$ and $\mathbf{x}^L$ are vectors with $K$ components:
\begin{eqnarray}
 \mathbf{x}^R &= \{ x_1^R, x_2^R, \cdots, x_K^R\}, \\
 \mathbf{x}^L &= \{ x_1^L, x_2^L, \cdots, x_K^L\}.
\end{eqnarray}
A set of finite dimensional equations for parameters $\mathbf{x}^R$ and $\mathbf{x}^L$ is obtained 
by the functional variation procedure.
Note that we set $\Phi(\mathbf{x}^L = 0 )$ to be consistent with the probabilistic interpretation,
so that
\begin{eqnarray}
\langle \Phi(\mathbf{x}^L = 0 ) | \Psi ( \mathbf{x}^R ) \rangle = 1,
\end{eqnarray}
which is the normalization condition for the probability distribution.
We, therefore, obtain the following equation which stems from an extremum of the action:
\begin{eqnarray}
\fl
\left[
\sum_{l=1}^K \left\langle \frac{\partial \Phi}{\partial x_{m}^L} \right|
 \left. \frac{\partial \Psi}{\partial x_{l}^R} \right\rangle
\frac{\rmd x_{l}^R}{\rmd t}
- \left\langle \left. \left. \frac{\partial \Phi}{\partial x_{m}^L} \right| \Omega
\right| \Psi \right\rangle 
\right]_{x_{m}^L = 0} = 0, \quad \mathrm{for} \,\, m = 1,2,\cdots, K.
\label{eq_variational_equation}
\end{eqnarray}

The only remaining procedure is to give an explicit ansatz for $\langle \Phi |$
and $| \Psi \rangle$.
This corresponds to the fact that we restrict a probability distribution 
$P_{\alpha 1}(n_\alpha, t)$ (or $P_{\alpha 0}(n_\alpha, t)$)
to a specific form with a few free parameters.
Although it is difficult to calculate the time evolution of the probability distribution directly,
the variational scheme enables us to get a set of time evolution equations
for the time-dependent parameters $\mathbf{x}^R$;
the time evolution equations for the parameters are determined variationally
through equation~\eref{eq_variational_equation}.
Note that in the variational scheme, it is necessary to set an adequate ansatz 
for $\langle \Phi |$ and $| \Psi \rangle$ 
in order to get qualitatively or quantitatively correct results.

\subsection{Poisson ansatz}

As for the choice of the ansatz in equation~\eref{eq_variational_equation},
only the Poisson ansatz has been proposed so far \cite{Sasai2003,Kim2007}.
The Poisson ansatz is a reasonable choice
because the steady-state probability distribution for a simple birth-death problem
is the Poisson distribution.
Furthermore, the Poisson ansatz is based on the coherent state,
which makes it easy to perform the variational calculation.

In the Poisson ansatz,
we assume the following ket vector
\begin{eqnarray}
| \psi_\alpha \rangle = 
\left( \begin{array}{c}
C_{\alpha 1} \exp\left[ X_{\alpha 1} ( a_\alpha^\dagger -1 )  \right] | 0_\alpha \rangle \\
C_{\alpha 0} \exp\left[ X_{\alpha 0} ( a_\alpha^\dagger -1 )  \right] | 0_\alpha \rangle 
\end{array} \right),
\end{eqnarray}
and as the bra ansatz, 
\begin{eqnarray}
\fl
\langle \phi_\alpha |
= \left( \begin{array}{cc}
\langle 0_\alpha | \exp\left( a_\alpha + \lambda_{\alpha 1}^{(0)} + \lambda_{\alpha 1}^{(1)} a_\alpha \right)
& \langle 0_\alpha | \exp\left( a_\alpha + \lambda_{\alpha 0}^{(0)} + \lambda_{\alpha 0}^{(1)} a_\alpha  \right)
\end{array} \right).
\label{eq_Poisson_bra_ansatz}
\end{eqnarray}
Note that although one might construct a bra ansatz with a lot of free parameters,
the same number of free parameters for the bra ansatz as that of the ket ansatz is enough 
for constructing the time evolution equations for the free parameters in the ket ansatz.
We therefore have in total $16$ parameters in the bra and ket variational functions;
\begin{eqnarray}
 \mathbf{x}^R &= \{ C_{A1}, C_{A0}, X_{A1}, X_{A0}, C_{B1}, C_{B0}, X_{B1}, X_{B0}    \}, \\
 \mathbf{x}^L &= \{ 
\lambda_{A1}^{(0)}, \lambda_{A0}^{(0)},
\lambda_{A1}^{(1)}, \lambda_{A0}^{(1)},
\lambda_{B1}^{(0)}, \lambda_{B0}^{(0)}, 
\lambda_{B1}^{(1)}, \lambda_{B0}^{(1)} \}.
\end{eqnarray}
Performing the variational calculation of equation~\eref{eq_variational_equation},
we finally have six coupled ordinary differential equations \cite{Kim2007};
the number of parameters for the ket ansatz is eight
but there are two constraints from the normalization of the probability:
$C_{A1} + C_{A0} = 1$ and $C_{B1} + C_{B0} = 1$.
In addition, all parameters in the bra ansatz are set to be zero finally,
and therefore there are only six equations.

\section{Beyond the Poisson ansatz}

Although it has been shown that the Poisson ansatz gives qualitatively appropriate results 
for the gene regulatory network \cite{Kim2007,Kim2007a},
the solution of the Poisson ansatz is not quantitatively correct.
Hence, it is necessary to develop the variational scheme beyond the Poisson ansatz.

In general, a state in the field theoretic description is described by 
$\sum_{n=0}^\infty P(n) | n \rangle$,
where $P(n)$ is a discrete probability distribution.
Note that $P(n)$ must be a discrete probability distribution because $n$ takes an integer value.
When we use the Poisson distribution as the probability $P(n)$,
we have the coherent states and then it is easy to calculate the functional variation.
However, for the other discrete probability distribution,
it is difficult to calculate the functional variation in equation~\eref{eq_variational_equation}.

In order to overcome the problems, we here propose a new ansatz for the variational scheme.
The new ansatz is based on the idea in which we use the superposition of the coherent states.
For example, when we want to have two parameters for the variational function,
the following ansatz for the ket state should be used:
\begin{eqnarray}
| \psi_\alpha \rangle = \left( \begin{array}{c}
C_{\alpha 1} \int_0^\infty \rmd x F(x; \mu_{\alpha 1}^{(1)}, \mu_{\alpha 1}^{(2)} )
\exp[ x (a_\alpha^\dagger -1) ] | 0_\alpha \rangle \\
C_{\alpha 0} \int_0^\infty \rmd x F(x; \mu_{\alpha 0}^{(1)}, \mu_{\alpha 0}^{(2)} )
\exp[ x (a_\alpha^\dagger -1) ] | 0_\alpha \rangle 
\end{array} \right) .
\label{eq_new_ansatz}
\end{eqnarray}
The new ansatz, \textit{the superposition ansatz}, means that
we take a superposition of the Poisson distributions with different mean values.
The `continuous' variational function $F(x)$ is a probability density with two parameters.
In the gene regulatory networks, the state $| n \rangle$ does not have negative $n$,
so the integral range of $F(x)$ should be taken as $x \geq 0$.
We note that the formalism can be extended to the case with more complicated variational function with many parameters.
The ansatz with only two free parameters in equation~\eref{eq_new_ansatz} is a simple case
beyond the Poisson ansatz.

Using the superposition ansatz,
we can easily perform the variational calculation because the variational functions are based on the coherent states.
In addition, the superposition ansatz enables us to use a continuous variational function.
Unlike using continuous approximations of master equations, such as the Kramers-Moyal expansion
and the system size expansion \cite{Risken1984,Gardiner2004},
the use of the continuous variational function in the superposition ansatz includes the
discrete characteristics of the original master equation due to the use of the coherent states.

As the bra ansatz, we here simply take
\begin{eqnarray}
\langle \phi_\alpha |
= \left( \begin{array}{c}
\langle 0_\alpha | \exp\left( a_\alpha + \lambda_{\alpha 1}^{(0)} 
+ \lambda_{\alpha 1}^{(1)} a_\alpha  + \lambda_{\alpha 1}^{(2)} (a_\alpha)^2 \right) \\
\langle 0_\alpha | \exp\left( a_\alpha + \lambda_{\alpha 1}^{(0)} 
+ \lambda_{\alpha 0}^{(1)} a_\alpha  + \lambda_{\alpha 0}^{(2)} (a_\alpha)^2 \right)
\end{array} \right)^T, 
\end{eqnarray}
where $T$ represents the transposed matrix.
Finally, we have the following $24$ parameters for the variational calculation
\begin{eqnarray}
 \mathbf{x}^R = \{ 
C_{A1}, C_{A0}, \mu_{A1}^{(1)},  \mu_{A0}^{(1)}, \mu_{A1}^{(2)}, \mu_{A0}^{(2)},
C_{B1}, C_{B0}, \mu_{B1}^{(1)},  \mu_{B0}^{(1)}, \mu_{B1}^{(2)}, \mu_{B0}^{(2)}
\} ,
\end{eqnarray}
\begin{eqnarray}
 \mathbf{x}^L = \{ 
\lambda_{A1}^{(0)}, \lambda_{A0}^{(0)}, 
\lambda_{A1}^{(1)}, \lambda_{A0}^{(1)},
\lambda_{A1}^{(2)}, \lambda_{A0}^{(2)},
\lambda_{B1}^{(0)}, \lambda_{B0}^{(0)}, 
\lambda_{B1}^{(1)}, \lambda_{B0}^{(1)} 
\lambda_{B1}^{(2)}, \lambda_{B0}^{(2)} 
\}.
\end{eqnarray}
Using the superposition ansatz of equation~\eref{eq_new_ansatz},
we have $10$ ordinary differential equations to be solved by numerical integration.
(The ket ansatz has $12$ parameters, but there are two constraints related to the normalization
of the probability, so that we have only $10$ equations.)

In what follows, we check the superposition ansatz by numerical experiments.
As the variational function with two parameters, we here take a gamma distribution;
\begin{eqnarray}
F(x; k, \theta) = x^{k-1} \frac{\exp(-x/ \theta)}{ \Gamma(k) \theta^k}.
\label{eq_gamma_ansatz}
\end{eqnarray}
The gamma function has the mean $k\theta$ and the variance $k \theta^2$.
As in the case of the Poisson ansatz,
a set of ordinary differential equations for the parameters 
are obtained by using a simple symbolic algebraic calculation in the field theoretic description.
The resulting equations are a little long,
so we write the resulting equations in the appendix.

\begin{figure}
\begin{center}
  \includegraphics[width=12cm,keepaspectratio,clip]{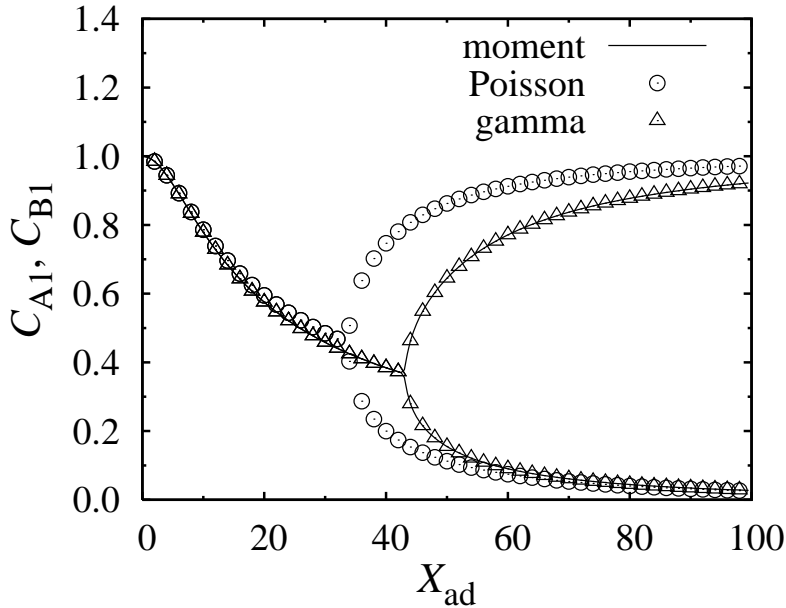} 
\caption{
Probabilities $C_{A1}$ and $C_{B1}$ in the long time limit (in the steady state).
The horizontal axis means the rescaled parameter $X_\textrm{ad} = (g_{1} + g_{0}) / (2 k_A)$.
At a certain critical point, there is the bifurcation from the monostable state to the bistable state.
The values of $C_{A1}$ and $C_{B1}$ are represented by the same symbol for simplicity.
We note that $C_{A1}$ and $C_{B1}$ take different stable states to each other in the bistable state.
The initial state determines which state ($C_{A1}$ or $C_{B1}$) takes the higher value in the bistable state.
}
\label{fig_2}
\end{center}
\end{figure}

We performed a numerical experiment in order to confirm the improvement achieved by the superposition ansatz.
We fixed all parameters except the protein synthesis rate $g_1 \equiv g_{A1} = g_{B1}$;
$k_A = k_B = 1, f_A = f_B = 0.5, h_A = h_B = f_A/500$, and $g_0 \equiv g_{A0} = g_{B0} = 0$,
which are the same parameters as in \cite{Kim2007}.
For various initial states of the variational parameters,
the steady state is obtained in the long time limit. 
Figure~\ref{fig_2} shows the probabilities $C_{A1}$ and $C_{B1}$ with which genes $A$ and $B$ are
in the active state, as a function of $X_\textrm{ad} = (g_{1} + g_{0})/(2k_A)$.
The values of $C_{A1}$ and $C_{B1}$ are represented by the same symbol for simplicity.
As shown in figure~\ref{fig_2}, the bifurcation from the monostable state to the bistable state is observed
as increasing $X_\textrm{ad}$.
We note that in the monostable state the values of $C_{A1}$ and $C_{B1}$ are the same,
but $C_{A1}$ and $C_{B1}$ take different stable states to each other in the bistable state.
It depends on the initial parameters which probability, $C_{A1}$ or $C_{B1}$, is larger than the other
in the bistable state.

The solid line in figure~\ref{fig_2} is obtained from the moment equations in \cite{Kim2007},
which is a closed set of equations and gives exact solutions for the present case.
The Poisson ansatz gives a qualitatively good results; the bifurcation is observed.
However, the bifurcation point is different from the result from the moment equations.
In contrast, the results from the gamma distribution ansatz are in quantitatively good agreement with 
the moment equations.
The numerical results confirm the validity of the superposition ansatz.

In the moment equations in reference~\cite{Kim2007},
the first and second moments of the protein number need to be taken into account.
In the Poisson ansatz, the mean and the variance should be same, and then
the Poisson ansatz does not give the quantitatively correct results
because the second moment depends on the first moment.
In contrast, the gamma distribution ansatz includes two free parameters
so the second moment of $x$ is independent of the first moment.
We consider that this is why the gamma distribution ansatz gives quantitatively correct results.
In addition, we have checked that the other ansatz, e.g., a log normal distribution, also works well
for calculating the bifurcation point correctly.
However, we note that there may be a suitable variational function for investigating higher correlations
in the gene regulatory networks.
Study of the applicability of the variational scheme will be important in the future.

\section{Concluding remarks}

In the present paper, a new ansatz for the variational scheme was proposed.
The superposition ansatz is based on the coherent states, 
so it gives us a straightforward extension of the variational scheme with the Poisson ansatz.
In addition, it enables us to use various continuous probability densities as the variational function.
The availability of the superposition ansatz was confirmed in a simple gene regulatory network.
The superposition ansatz gives a quantitatively correct solution, while the Poisson ansatz
is adequate only qualitatively.

The concept of the superposition of the Poisson distributions
seems to be related to the Poisson representation \cite{Gardiner2004}.
The coefficient function in the Poisson representation can take complex numbers,
so that it is not always true that the coefficient function corresponds to the probability distribution.
The relationship between the Poisson representation and the quantum field theoretic representation
has been pointed out\cite{Droz1994},
and actually, our variational scheme is related to the Poisson representation;
it is easy to see that the superposition ansatz restricts the coefficient function in the Poisson representation 
to being a certain variational function.
This correspondence between the superposition ansatz and the Poisson representation
would give us further extensions of the superposition ansatz;
it might be possible to use a function of complex variable as the variational function.
This is a future work.

The variational method and the quantum field theoretical description would give
new and useful schemes of approximation for complicated master equations.
For example, the superposition ansatz enables us to extend the variational scheme 
to multivariate cases \cite{Ohkubo_pre}.
These approximation methods are important for researching complex systems such as biological systems
and social systems.
Furthermore, it may be possible to study the complex systems more analytically 
by using the quantum field theoretical description.
Such researches would give deep insight into the complex systems.


\appendix 
\section{Time evolution equations in the superposition Ansatz}

From equation~\eref{eq_variational_equation} and the superposition ansatz of \eref{eq_new_ansatz},
a set of coupled ordinary differential equations are derived.
Here, we use the following notation for simplicity:
$F(x; \mu_{\alpha 1}^{(1)}, \mu_{\alpha 1}^{(2)} ) \equiv F_{\alpha 1}(x)$.
Performing the variational calculation, we obtain the following five time-evolution equations 
for the parameters related to gene $A$:
\begin{eqnarray}
\fl
\frac{\rmd C_{A1}}{\rmd t} = -C_{A1} \left( 
C_{B1} \frac{h_A}{2} \int_0^\infty \rmd x x^2 F_{B1}(x) + C_{B0} \frac{h_A}{2} \int_0^\infty \rmd x x^2 F_{B0}(x)
\right) + f_A C_{A0}
\label{appendix_1},
\end{eqnarray}
\begin{eqnarray}
\fl
\frac{\rmd C_{A1}}{\rmd t} \int_0^\infty \rmd x x F_{A1}(x)
+ C_{A1} \frac{\rmd \mu_{A1}^{(1)}}{\rmd t} \int_0^\infty \rmd x x \frac{\partial F_{A1}(x)}{\partial \mu_{A1}^{(1)}}
+ C_{A1} \frac{\rmd \mu_{A1}^{(2)}}{\rmd t} \int_0^\infty \rmd x x \frac{\partial F_{A1}(x)}{\partial \mu_{A1}^{(2)}}
\nonumber \\
= C_{A1}\left[ g_{A1} -k  \int_0^\infty \rmd x F_{A1}(x) \right]
+ C_{A0} f_A \int_0^\infty \rmd x x F_{A0}(x) \nonumber \\
- \frac{h_A}{2} C_{A1} \int_0^\infty \rmd x_A  x_A F_{A1}(x_A) \nonumber \\
\times \left\{ 
C_{B1} \int_0^\infty \rmd x_B  x_B^2 F_{B1}(x_B)
+ C_{B0} \int_0^\infty \rmd x_B  x_B^2 F_{B0}(x_B) 
\right\} 
\label{appendix_2},
\end{eqnarray}
\begin{eqnarray}
\fl
\frac{\rmd C_{A0}}{\rmd t} \int_0^\infty \rmd x x F_{A0}(x)
+ C_{A0} \frac{\rmd \mu_{A0}^{(1)}}{\rmd t} \int_0^\infty \rmd x x \frac{\partial F_{A0}(x)}{\partial \mu_{A0}^{(1)}}
+ C_{A0} \frac{\rmd \mu_{A0}^{(2)}}{\rmd t} \int_0^\infty \rmd x x \frac{\partial F_{A0}(x)}{\partial \mu_{A0}^{(2)}}
\nonumber \\
= C_{A0}\left[ g_{A0} -k  \int_0^\infty \rmd x F_{A0}(x) \right]
- C_{A0} f_A \int_0^\infty \rmd x x F_{A0}(x) \nonumber \\
+ \frac{h_A}{2} C_{A1} \int_0^\infty \rmd x_A  x_A F_{A1}(x_A) \nonumber \\
\times \left\{ 
C_{B1} \int_0^\infty \rmd x_B  x_B^2 F_{B1}(x_B)
+ C_{B0} \int_0^\infty \rmd x_B  x_B^2 F_{B0}(x_B) 
\right\} 
\label{appendix_3},
\end{eqnarray}
\begin{eqnarray}
\fl
\frac{\rmd C_{A1}}{\rmd t} \int_0^\infty \rmd x x^2 F_{A1}(x)
+ C_{A1} \frac{\rmd \mu_{A1}^{(1)}}{\rmd t} \int_0^\infty \rmd x x^2 \frac{\partial F_{A1}(x)}{\partial \mu_{A1}^{(1)}}
+ C_{A1} \frac{\rmd \mu_{A1}^{(2)}}{\rmd t} \int_0^\infty \rmd x x^2 \frac{\partial F_{A1}(x)}{\partial \mu_{A1}^{(2)}}
\nonumber \\
= C_{A1} \int_0^\infty \rmd x F_{A1}(x) \left[ 2 g_{A1}x - 2k x^2 \right]
+ C_{A0} f_A \int_0^\infty \rmd x x^2 F_{A0}(x) \nonumber \\
- \frac{h_A}{2} C_{A1} \int_0^\infty \rmd x_A  x_A^2 F_{A1}(x_A) \nonumber \\
\times \left\{ 
C_{B1} \int_0^\infty \rmd x_B  x_B^2 F_{B1}(x_B)
+ C_{B0} \int_0^\infty \rmd x_B  x_B^2 F_{B0}(x_B) 
\right\}
\label{appendix_4},
\end{eqnarray}
\begin{eqnarray}
\fl
\frac{\rmd C_{A0}}{\rmd t} \int_0^\infty \rmd x x^2 F_{A0}(x)
+ C_{A0} \frac{\rmd \mu_{A0}^{(1)}}{\rmd t} \int_0^\infty \rmd x x^2 \frac{\partial F_{A0}(x)}{\partial \mu_{A0}^{(1)}}
+ C_{A0} \frac{\rmd \mu_{A0}^{(2)}}{\rmd t} \int_0^\infty \rmd x x^2 \frac{\partial F_{A0}(x)}{\partial \mu_{A0}^{(2)}}
\nonumber \\
= C_{A0} \int_0^\infty \rmd x F_{A0}(x) \left[ 2 g_{A0}x - 2k x^2 \right]
- C_{A0} f_A \int_0^\infty \rmd x x^2 F_{A0}(x) \nonumber \\
+ \frac{h_A}{2} C_{A1} \int_0^\infty \rmd x_A  x_A^2 F_{A1}(x_A) \nonumber \\
\times \left\{ 
C_{B1} \int_0^\infty \rmd x_B  x_B^2 F_{B1}(x_B)
+ C_{B0} \int_0^\infty \rmd x_B  x_B^2 F_{B0}(x_B) 
\right\}.
\label{appendix_5}
\end{eqnarray}
We have similar five equations for gene $B$,
which are expressed by the exchange of the indexes ($A \leftrightarrow B$)
for equations~\eref{appendix_1} $\sim$ \eref{appendix_5}.
We note that there are restrictions for the normalization of probability
$C_{\alpha 0} = 1 - C_{\alpha 1}$.

When we use the gamma distribution \eref{eq_gamma_ansatz} for the superposition ansatz,
the integral factors in equations~\eref{appendix_1} $\sim$ \eref{appendix_5}
are simply replaced by 
\begin{eqnarray}
\int_0^\infty \rmd x x \frac{\partial F_{\alpha i}(x)}{\partial \mu_{\alpha i}^{(1)}} 
= \frac{\partial}{\partial \mu_{\alpha i}^{(1)}} \int_0^\infty \rmd x x  F_{\alpha i}(x)
= \mu_{\alpha i}^{(2)},
\end{eqnarray}
\begin{eqnarray}
\int_0^\infty \rmd x x^2 \frac{\partial F_{\alpha i}(x)}{\partial \mu_{\alpha i}^{(1)}} = 
( \mu_{\alpha i}^{(2)} )^2 + 2 \mu_{\alpha i}^{(1)} ( \mu_{\alpha i}^{(2)} )^2,
\end{eqnarray}
\begin{eqnarray}
\int_0^\infty \rmd x x \frac{\partial F_{\alpha i}(x)}{\partial \mu_{\alpha i}^{(2)}} = \mu_{\alpha i}^{(1)},
\end{eqnarray}
\begin{eqnarray}
\int_0^\infty \rmd x x^2 \frac{\partial F_{\alpha i}(x)}{\partial \mu_{\alpha i}^{(2)}} = 
2 \mu_{\alpha i}^{(2)} ( \mu_{\alpha i}^{(1)}  + ( \mu_{\alpha i}^{(1)} )^2 ),
\end{eqnarray}
where 
$\alpha \in \{ A, B \}$
and $i \in \{ 0, 1\} $.

In order to evaluate the time evolution of the parameters related to gene A numerically,
we need to calculate $dC_{A1}/dt$, $dC_{A0}/dt$, 
$d\mu_{A1}^{(1)}/dt$, $d\mu_{A0}^{(1)}/dt$, $d\mu_{A1}^{(2)}/dt$, and $d\mu_{A0}^{(2)}/dt$.
From equation~\eref{appendix_1}, we have $dC_{A1}/dt$,
and then $dC_{A0}/dt$ is calculated using
\begin{eqnarray}
\frac{dC_{A0}}{dt} = - \frac{dC_{A1}}{dt}.
\end{eqnarray}
Because equations~\eref{appendix_2} and \eref{appendix_4}
are linear simultaneous equations in $d\mu_{A1}^{(1)}/dt$ and $d\mu_{A1}^{(2)}/dt$,
it is easy to calculate $d\mu_{A1}^{(1)}/dt$ and $d\mu_{A1}^{(2)}/dt$.
$d\mu_{A0}^{(1)}/dt$ and $d\mu_{A0}^{(2)}/dt$
are also calculated from linear simultaneous equations~\eref{appendix_3} and \eref{appendix_5}.
For the time evolution of the parameters related to gene B, we perform the same procedures.

\end{document}